\documentclass[12pt]{article}
\usepackage{graphicx}

\def \azeL{{A_0^L}}
\def \azeR{{A_0^R}}
\def \apaL{{A_\parallel^L}}
\def \apaR{{A_\parallel^R}}
\def \apeL{{A_\perp^L}}
\def \apeR{{A_\perp^R}}

\def \be{\begin{equation}}
\def \ee{\end{equation}}
\def \bea{\begin{eqnarray}}
\def \eea{\end{eqnarray}}
\def \ben{\begin{enumerate}}
\def \een{\end{enumerate}}
\def \bit{\begin{itemize}}
\def \eit{\end{itemize}}
\def \Bbar{\overline{\kern -0.24em B}}

\def \del#1{\Delta^{#1}_{\tilde D_L}}

\def \Im{{\rm{Im}}}
\def \Re{{\rm{Re}}}

\def \Ceff{ C_9^{\rm{eff}}}

\newcommand\pubnumber{UAB-FT-686, MZ-TH/10-48}
\newcommand\pubdate{\today}

\def\Title#1{\begin{center} {\Large #1 } \end{center}}
\def\Author#1{\begin{center}{ \sc #1} \end{center}}
\def\Address#1{\begin{center}{ \it #1} \end{center}}

\newcommand\pubblock{\rightline{\begin{tabular}{l} \pubnumber\\
         \pubdate  \end{tabular}}}
\newenvironment{Abstract}{\begin{quotation}  }{\end{quotation}}
\newenvironment{Presented}{\begin{quotation} \begin{center} 
             PRESENTED AT\end{center}\bigskip 
      \begin{center}\begin{large}}{\end{large}\end{center} \end{quotation}}

\def\beq{\begin{equation}}
\def\eeq#1{\label{#1}\end{equation}}
\def\eeqn{\end{equation}}

\def\beqa{\begin{eqnarray}}
\def\eeqa#1{\label{#1}\end{eqnarray}}
\def\eeqan{\end{eqnarray}}

\let\bar=\overbar

\def\Dslash{\not{\hbox{\kern-4pt $D$}}}
\def\dslash{\not{\hbox{\kern-2pt $\del$}}}

\def\ee{e^+e^-}

\def\msb{{\bar{\ssstyle M \kern -1pt S}}}

\begin{document}
\begin{titlepage}
\pubblock

\vfill
\Title{Symmetries in the angular  distribution\\  of exclusive  semileptonic $B$ decays} 
\vfill
\Author{Ulrik Egede} 
\Address{Imperial College London, London SW7~2AZ, United Kingdom}
\Author{Tobias Hurth}
\Address{	Institute for Physics, Johannes Gutenberg-University, D-55099 Mainz, Germany}
\Author{Joaquim Matias}
\Address{Universitat Aut\`onoma de Barcelona, 08193 Bellaterra, 
Barcelona, Spain}
\Author{Marc Ramon}
\Address{	Universitat Aut\`onoma de Barcelona, 08193 Bellaterra, 
Barcelona, Spain}
\Author{Will Reece}
\Address{	CERN, Dept. of Physics, CH-1211 Geneva 23, Switzerland}

\vfill
\begin{Abstract}

We discuss a method to construct observables protected against QCD uncertainties based on the angular distribution of the exclusive  $B_d \to K^{*0}(\to K\pi) l^+ l^-$ decay. We focus on the identification and the interpretation of all the symmetries of the  distribution. They constitute a key ingredient to construct a set of so-called transverse observables. We work in the framework of QCD factorization at NLO  supplemented by an estimate of  power-suppressed $\Lambda/m_b$ corrections. A discussion of the new physics properties of two of the transverse asymmetries,  $A_T^{(2)}$ and $A_T^{(5)}$,  is presented. A comparison between the transverse asymmetry $A_T^{(2)}$ and the forward-backward asymmetry shows that $A_T^{(2)}$ emerges as an improved version of it.

\end{Abstract}
\vfill
\begin{Presented}
the 6th International Workshop on the CKM Unitarity\\ Triangle,
 University of Warwick, UK, 6.-10.9.2010 by J.M\\  and \\
the 35th International Conference on High Energy Physics\\ ICHEP2010,
Paris, France, 21.-28.7.2010  by T.H.
\end{Presented}
\vfill
\end{titlepage}
\def\thefootnote{\fnsymbol{footnote}}
\setcounter{footnote}{0}



\noindent The next decade will be hopefully  dominated by the new physics discoveries at LHC. In this scenario flavour physics will play an important and complementary role  to direct discovery, by exploring the flavour sector of the theory that lies beyond the SM. Rare B decays are  sensitive probes to new physics signals (for a recent review see Ref. \cite{Hurth:2010tk}). Amongst them the semileptonic exclusive decay $B \to K^* l^+ l^-$ is particularly important due to its  very rich phenomenology. Some of the observables constructed out of this decay are:  the forward-backward asymmetry, $A_{FB}$, and its zero 
\cite{Burdman:1998mk,Beneke:2001at,Alok:2009tz}, the isospin asymmetry,  $A_I$,  
\cite{Feldmann:2002iw,Beneke:2004dp}, and  the transverse 
observables $A_T^{(i)}$ (i=2,3,4,5)  based on the four-body angular distribution when the $K^*$ decays into a $K\pi$ pair \cite{Kruger:2005ep,Lunghi:2006hc,Egede:2008uy,Egede:2010zc}. 
 Also the coefficients of the angular distribution \cite{Altmannshofer:2008dz} or ratios between different $q^2$ regions \cite{Lunghi:2010tr} are used to define observables.
The main focus of this paper is to provide a guideline for the construction of transverse observables ($A_T^{(i)}$). 
These observables maximize the sensitivity to new physics  and, at the same time, exhibit a minimal hadronic uncertainty, in particular, to the poorly known soft form factors.

\section{General method}

In this section we will describe  the basis of the method recently completed in \cite{Egede:2010zc} to construct robust transverse\footnote{Indeed only $A_T^{(2)}$ and $A_T^{(5)}$ are strictly transverse observables, $A_T^{(3)}$ and $A_T^{(4)}$ being also sensitive to the longitudinal spin amplitudes should, for consistency, be called transverse/longitudinal observables.} observables.
 The method  is sufficiently general to be applied to angular distributions with similar properties.
The steps of the method are: 1) use the helicity amplitudes of the $K^*$ as the key ingredients to construct a quantity where the soft form factor dependence cancels at LO (amplitudes in the large recoil limit are very useful to check this cancellation)  2) identify all symmetries of the distribution with respect to transformations of the $K^*$ spin amplitudes 3) check that the constructed quantity fulfils all the symmetries to identify it as an observable 4) express the observable in terms of the coefficients of the distribution.  As a by-product of the method hidden correlations between the coefficients of the distribution may arise. These correlations have proven to  be important for the stability of the fit and also     provide a powerful extra experimental check. 

Our main source of information is the differential decay distribution of the decay ${ {\bar B_d}\to {\bar K^{*0}}(\to K^-\pi^+)l^+l^-}$ with the $K^{*0}$ on the mass shell. This distribution is a function of four variables
\begin{equation}
\label{eq:differential decay rate}
  \frac{d^4\Gamma}{dq^2\, d\cos\theta_l\, d\cos\theta_{K}\, d\phi} =
   \frac{9}{32\pi} J(q^2, \theta_l, \theta_{K}, \phi)\nonumber
\end{equation}
where
 $ q^2=s$ is the square of the lepton-pair invariant mass, $ \theta_l$ is the angle between $\vec{p_{l^+}}$ in  $l^+l^-$ rest frame  and the di-lepton's direction in rest frame of $\bar B_d$, $\theta_K$ is the angle between $\vec{p_{K^-}}$ in ${\bar K^{*0}}$ rest frame and the direction of the $\bar K^{*0}$ in rest frame of $\bar B_d$, and finally $\phi$ is the angle between the di-lepton plane and the $K-\pi$ plane.
The function $J(q^2, \theta_l, \theta_K, \phi)$ splits into the following coefficients of the distribution \cite{Egede:2010zc,Altmannshofer:2008dz}
\begin{eqnarray} 
 &J(q^2, \theta_l, \theta_K, \phi) =\nonumber\\
   & J_{1s} \sin^2\theta_K + J_{1c} \cos^2\theta_K
     + (J_{2s} \sin^2\theta_K + J_{2c} \cos^2\theta_K) \cos 2\theta_l + J_3 \sin^2\theta_K \sin^2\theta_l \cos 2\phi 
\nonumber \\       
    & + J_4 \sin 2\theta_K \sin 2\theta_l \cos\phi  + J_5 \sin 2\theta_K \sin\theta_l \cos\phi+ (J_{6s} \sin^2\theta_K +  {J_{6c} \cos^2\theta_K})  \cos\theta_l 
\nonumber \\      
    & + J_7 \sin 2\theta_K \sin\theta_l \sin\phi  + J_8 \sin 2\theta_K \sin 2\theta_l \sin\phi + J_9 \sin^2\theta_K \sin^2\theta_l \sin 2\phi\,.\nonumber
\end{eqnarray}
These coefficients $J_i$ with $i=1s,1c,2s,2c,3-5,6s,6c,7-9$ are in turn functions of the  amplitudes $A^{(L,R)}_{\perp,\|,0,t,S}$ \cite{Egede:2010zc,Altmannshofer:2008dz} ($A_{\perp,\|,0}$ are linear combinations of the well-known helicity amplitudes $H_{+1,-1,0}$).  The counting of the coefficients of the angular distribution and of the theoretical spin amplitudes depends on whether scalar
interactions  are relevant in the analysis  or not.
If we include them we have 8 complex amplitudes ($A_{\perp,||,0,(L,R)S,t}$) and 12 experimental inputs ($J_i$), while if 
no scalar amplitudes are considered  we would have just 7 complex amplitudes ($A_{\perp,||,0,(L,R),t}$) and 11 experimental coefficients  ($J_{6c}=0$).  If we neglect the mass of the lepton in addition,  the number of complex spin amplitudes gets further reduced to 6  ($A_t=0$).  

\section{Symmetries of the distribution}

Experimental ($J_i$) and theoretical ($A_i$) degrees of freedom have to match. The equation that defines this matching is
$ n_C-n_d=2 n_A-n_s$,
where
  $n_C$ is the number of coefficients of the differential distribution ($J_i$), $ n_d$ is the number of relations between the $J_i$,  $n_A$ is the number of spin amplitudes, 
and $n_s$ is the number of symmetries of the distribution.

We will focus here on the case of massless leptons with no scalars. The parameters of the equation are then
 $n_C=11$, $n_d=3$ ($J_{1s}=3 J_{2s}$, $J_{1c}=-J_{2c}$, and a  third more complex relation), $n_A=6$ (spin amplitudes),  $n_s=4$ symmetries.
One of the main results in Ref. \cite{Egede:2010zc} was to identify the fourth and last symmetry (three of them were found in Ref. \cite{Egede:2008uy}). Moreover,  a non-trivial hidden correlation between the coefficients of the distribution was discovered.  

One important question arises at this point: how do we know that there are four symmetries without having found  first the new non-trivial hidden correlation?


In order to count the number of symmetries we define an 
{ infinitesimal symmetry} transformation of the distribution:
$
  \label{eq:InfinitesimalTransformationDef}
  \vec{A'} = \vec{A} + \delta \vec{S}\nonumber
$
%
where
\begin{eqnarray}
  \label{eq:Avector}
  \vec{A} & = & 
      \left(\Re(\apeL),\Im(\apeL),\Re(\apaL),\Im(\apaL),\Re(\azeL),\Im(\azeL), 
      \phantom{)}\right. \nonumber \\ 
      & &\left. \phantom{(}  \Re(\apeR),\Im(\apeR),\Re(\apaR),\Im(\apaR),\Re(\azeR),\Im(\azeR)\right) \,.\nonumber
\end{eqnarray}
%
%
%
$\vec{S}$ represents a symmetry of the distribution if and only 
if  $\forall i \in (J_{1s}...J_9):{\vec {\nabla}}(J_i)\perp\vec{S}$.
%

There are  as many independent infinitesimal symmetries  as linearly independent vectors $\vec{S}_j$, with $j=1,..n_s$ satisfying the above constraint. In the case of massless leptons with no scalars four of those vectors $\vec{S}_j$ were found   \cite{Egede:2010zc}.
This was the first proof that four and no more symmetries are present.




{
 The explicit form of the four continuous independent symmetry transformations\footnote{Sometimes it 
 might be  a non-trivial task to find a  continuous symmetry associated to an infinitesimal one.} of the amplitudes that leave the differential distribution invariant are \cite{Egede:2010zc}: 
  \begin{equation}
  \label{eq:SymMassless}
  n_i^{'} = 
  \left[
    \begin{array}{ll}
      e^{i\phi_L} & 0 \\
      0 & e^{-i \phi_R}
    \end{array}
  \right]
  \left[
    \begin{array}{rr}
      \cos \theta & -\sin \theta \\
      \sin \theta & \cos \theta
    \end{array}
  \right]
    \left[
    \begin{array}{rr}
      \cosh i \tilde{\theta} &  -\sinh i \tilde{\theta} \\
      -\sinh i \tilde{\theta} & \cosh i \tilde{\theta}
    \end{array}
  \right]
  n_i 
\end{equation}
 where we have defined   $n_1=(\apaL, \apaR^*)$, $n_2=(\apeL, - \apeR^*)$ and $n_3=(\azeL, \azeR^*)$. 
%
The first two symmetries (phase transformations) are a consequence of the freedom to pick up an arbritrary and  different global phase  for the L and R non-interfering amplitudes. The third and fourth symmetry corresponds to  the experimental impossibility to measure a simultaneous change of helicity and handedness of the current (a helicity +1 state with a left handed current transforms into a helicity -1 with a right handed current).

But, what have we learnt from using this symmetry approach? The answer to this question is twofold. On the one side, it basically gives freedom to construct an optimal observable out of the spin amplitudes. The symmetries allow to bypass  the strong restriction of taking each coefficient of the distribution as an observable and permits to construct the best, i.e. most sensitive to NP, combination of them. The only requirement to fulfil is that the constructed quantity  has to respect these symmetries   (in order to be promoted to an observable). On the other side, the symmetries of the distribution are necessary  to find a solution of the system of the spin amplitudes in terms of the coefficients of the distribution; in particular 
it allows us to identify new hidden correlations which turn out to be important for the stability of the 
experimental fit.

Indeed we found in Ref.\cite{Egede:2010zc} that  all the physical information of the distribution is encoded in the three  moduli and the
complex scalar products of the vectors $n_i$, 
{ \begin{eqnarray} \small \label{eq1} |n_1|^2=\frac{2}{3} J_{1s} - J_3 \,, \quad
  \quad |n_2|^2&=&\frac{2}{3} J_{1s} + J_3 \,, \quad \quad 
  |n_3|^2=J_{1c} \nonumber \\ 
  n_1 \cdot n_2=\frac{J_{6s}}{2}- i J_9 \,, \quad \quad 
  n_1 \cdot n_3&=&\sqrt{2} J_4 - i \frac{J_7}{\sqrt{2}} \,, \quad \quad
  \label{eq6}
  n_2 \cdot n_3=\frac{J_5}{\sqrt{2}} - i \sqrt{2} J_8 \,. \nonumber
\end{eqnarray}}
 The symmetries guarantee the invariance of these moduli and scalar products.}
Using the freedom given by the symmetries to fix certain parameters to zero, the system of $A$'s can be solved in terms of $J$'s. In particular, we choose the left global phase ($\phi_L$) such that ${\rm Im} A_{\|}^L=0$, the right global phase symmetry ($\phi_R$) such that ${\rm Im} A_{\|}^R=0$ (simplicity) and one of the continuous $L \leftrightarrow R$ rotation $\theta$ to fix ${\rm Re} A_{\|}^R=0$.
The system is then easily solved  as shown in Table 1. Still one last equation remains
%
{\begin{table}
 \begin{tabular}{ccc} 
    \hline 
    $A_\perp^L=\left[\frac{\frac{4}{9} 
J_{1s}^{2}-J_3^2 - \frac{1}{4} J_{6s}^{2} - J_9^2}{\frac{2}{3} J_{1s} - 
J_3}\right]^\frac{1}{2} e^{i \phi_\perp^L}$ & \quad \quad \quad  & $ A_\perp^R= - \frac{\left(J_{6s} - 
    2  i J_9 \right)}{2 \sqrt{\frac{2}{3} J_{1s} - J_3}},$  \\
      $A_\|^L=0$ & \quad \quad \quad  & $A_\|^R=\sqrt{\frac{2}{3} J_{1s} - J_3}$  \\
   $A_0^L=\left[
\frac{J_{1c} \left(\frac{2}{3} J_{1s} - J_3 \right) - 2 J_4^2- \frac{1}{2} 
J_7^2}{\frac{2}{3} J_{1s} -
J_3}\right]^\frac{1}{2} e^{i  \phi_0^L} $
 &  \quad \quad \quad & $A_0^R=  \frac{2 J_4 - i 
  J_7}{
  \sqrt{\frac{4}{3} J_{1s}-2 J_3}}$
\\ \hline \end{tabular}  \caption{Explicit solution of the spin amplitudes in terms of the coefficients of the distribution for the massless case without scalars.}\end{table}}
{\begin{eqnarray}\label{phases}
e^{i (\phi_{\perp}^{L}-\phi_{0}^{L})} = \frac{ J_5 \left(\frac{2}{3} J_{1s}-J_3 \right)-J_4 J_{6s} -J_7 
J_9 
-i \left(\frac{4}{3} J_{1s} J_8 - 2 J_3 J_8 + 2 J_4 J_9 - \frac{1}{2} 
J_{6s} J_7 \right) }{\left[
2 \left(\frac{4}{9} J_{1s}^{2} -J_3^2- \frac{1}{4} J_{6s}^{2}-J_9^2 
\right) \left(J_{1c} \left( \frac{2}{3} J_{1s} - J_3 \right) - 2 J_4^2 - 
\frac{1}{2} J_7^2 \right)\right]^{1/2}} \nonumber .
\end{eqnarray}}
This equation has two important consequences. First, it  represents another proof of the existence of the fourth symmetry manifesting itself in the freedom to choose either $\phi_{\bot}^L$ or $\phi_0^{L}$ = 0. And second, the condition of the LHS of this equation  
being a phase impose the following non-trivial constrain on the RHS:
\begin{eqnarray} \label{hidden}
J_{1c} =-J_{2c}= &  6\frac{ (2 J_{1s}+ 3 J_3) \left(4 J_4^2+J_7^2\right)
+ ( 2 J_{1s} - 3
J_3)
\left(J_5^2+4 J_8^2 \right)}{16 J_{1s}^{2} -
9
\left(4 J_3^2+ J_{6s}^{2} + 4 J_9^2 \right)} \nonumber \\
& \ -36 \frac{J_{6s} (J_4 J_5 +
 J_7 J_8) + J_9 (J_5 J_7 - 4 J_4 J_8)}{16 J_{1s}^{2} -
9
\left(4 J_3^2+ J_{6s}^{2} + 4 J_9^2 \right)}\equiv f \,. 
\end{eqnarray}
We emphasize that this equation holds in the case without scalar amplitudes and under the assumption that the mass of the leptons can be neglected. If  scalar
amplitudes are relevant  ($J_{1c} \neq -J_{2c}$) the equation $-J_{2c}=f$ is still fulfilled while  $J_{1c}\neq f$. 
Taking into account  the mass of the leptons,  one can derive a  similar expression only if there are 
no scalar amplitudes included in the analysis. 
As a consequence, if $J_{1c}=f$ is not fulfilled and large deviations are observed (small deviations may be due to the massive  terms)  this  would signal the presence of scalars. 
On the contrary, if the equation $-J_{2c}=f$ is not fulfilled and large deviations are observed it might point to an experimental problem.

\section{Construction of transverse observables: $A_T^{(i)}$}

Following the previous steps we constructed four different robust observables. Two of them $A_T^{(2)}$ and $A_T^{(5)}$ are only sensitive to the transverse amplitudes, while $A_T^{(3)}$ and $A_T^{(4)}$ also have sensitivity to the longitudinal spin amplitude. In this section we will focus on the properties of the former.  
The computation of spin amplitudes $A_{\perp,\|,0}^{(L,R)}$ is done at the NLO level within  the framework of QCD-factorization  \cite{Beneke:2001at} . They are functions of the long-distance $B \to K^*$ form factors ($ A_{0,1,2}(q^2), V(q^2),  T_{1,2,3}(q^2)$)(see \cite{Khodjamirian:2010vf} for a recent update) and  of the  short-distance Wilson coefficients  ($ C^{\rm eff}_{7},  C^{\rm eff\prime}_ {7}, { \Ceff},{ C_{10}}, {\Ceff}^\prime, C_{10}^\prime$) (for precise definitions see Refs. \cite{Egede:2008uy,Egede:2010zc}). 
In the heavy quark and large $E_{K^*}$ limit all form factors can be expressed in terms of just two soft form factors $\xi_\perp(E_K^*)$ and $\xi_\|(E_K^*)$ \cite{Charles}.
However these relations receive two types of  corrections: order $\alpha_s$ \cite{Beneke:2001at,Beneke:2004dp}(coming from NLO-QCDf) and power suppressed  $\Lambda/m_b$ corrections estimated to be of ${\cal O}(10\% )$.
Both were included in our computation of the spin amplitudes at the NLO level in Refs. \cite{Egede:2008uy,Egede:2010zc}.
\begin{figure}[htb]
\includegraphics[height=.33\textwidth,width=.33\textwidth]{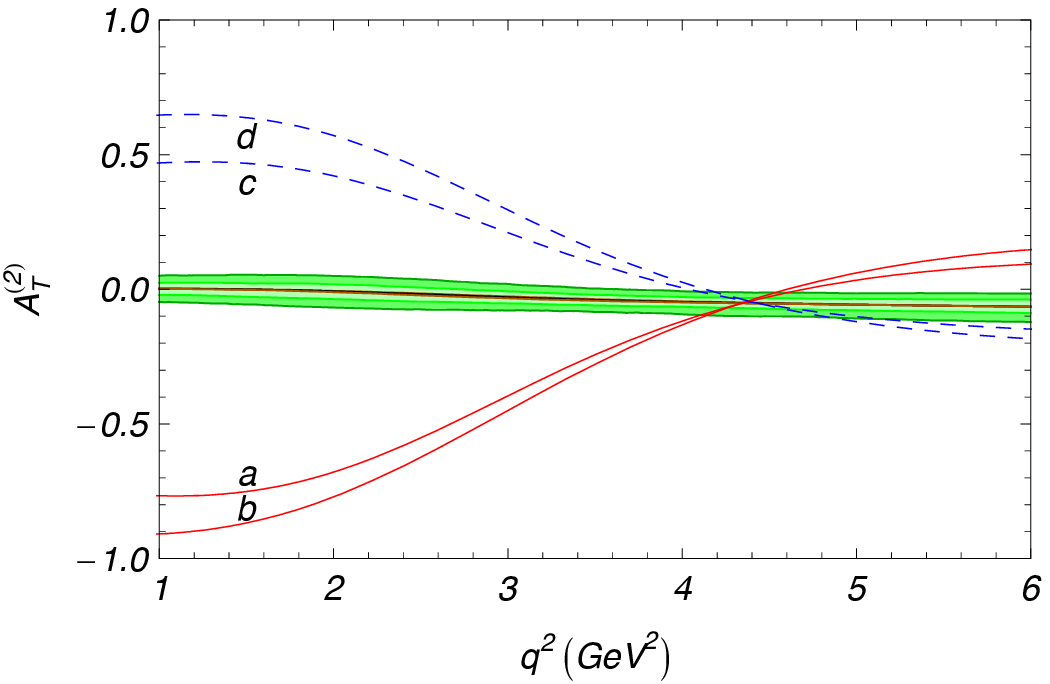}\includegraphics[height=.33\textwidth,width=.33\textwidth]{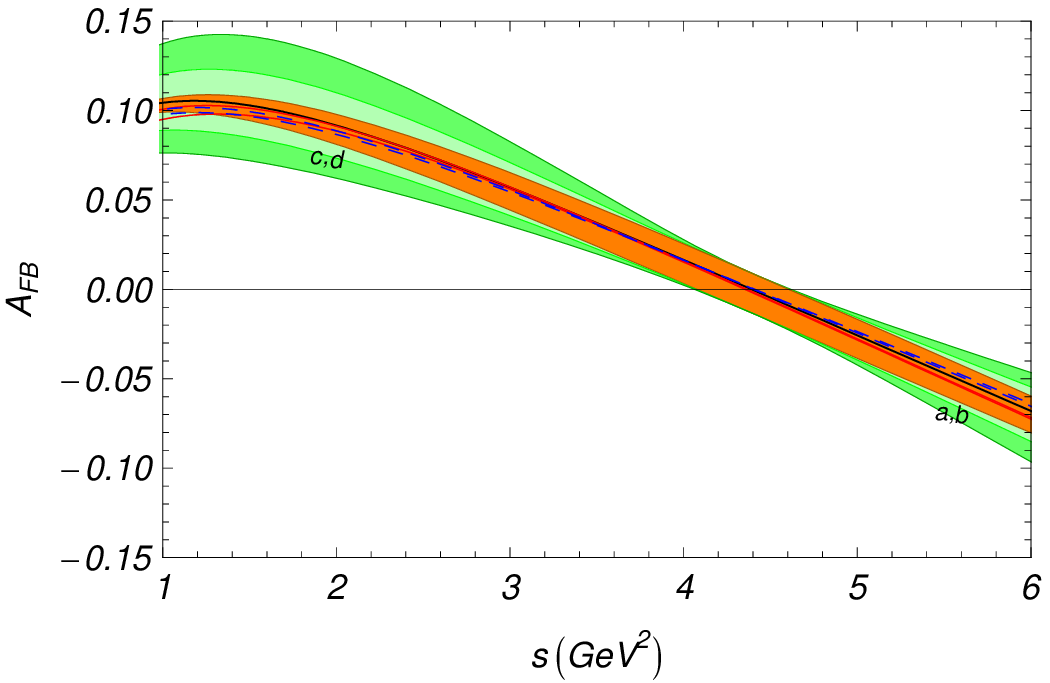}\includegraphics[height=.33\textwidth,width=.33\textwidth]{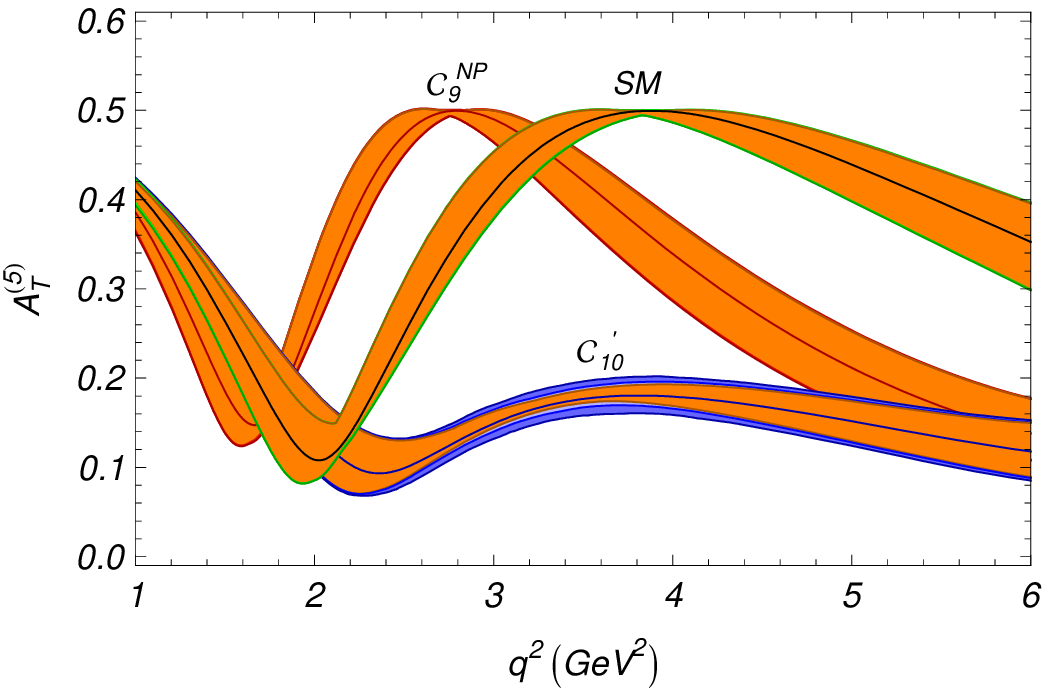}
\caption{Left: $A_T^2$ in SM (green band) with four NP benchmarks (\cite{Egede:2010zc}). Center: $A_{FB}$ for the same cases. Right: $A_T^5$ in the SM and for different values of $C_9^{eff}$ and $C_{10}^\prime$ 
(for more details see Ref.\cite{Egede:2010zc}).}
\label{fig:magnet}
\end{figure}

The  observable $A_T^{(2)}$ 
was first  proposed in Ref. \cite{Kruger:2005ep}:
\begin{equation}
    \label{eq:AT2Def}
    A_T^{(2)} =\frac{|A_\perp|^2 - |A_\||^2}{|A_\perp|^2 + |A_\||^2} \,. 
    \nonumber  \end{equation}
It is built to signal deviations from the left-handed structure of the SM: $A_T^2|_{SM}\sim 0$. 
We restricted our analysis  to the $1<q^2<6 \,{\rm GeV}^2$ region (its extension to $q^2>14\, {\rm GeV}^2$ was described in \cite{Bobeth:2010wg}). Some of the most important properties are:
\begin{itemize}
\item The soft form factor dependence  cancels exactly at LO and a very mild dependence at NLO is observed.
\item In the large recoil limit for the spin amplitudes  $A_T^{(2)}$ simplifies to
 \begin{equation}
    \label{eq:AT2Def} 
    A_T^{(2)} 
\sim 4 {C_7^{\rm eff}}^\prime \frac{m_b M_B}{s} \frac{\Delta_{-}+\Delta_{+}^*}{2 C_{10}^2+|\Delta_{-}|^2+|\Delta_{+}|^2}  \, \nonumber
  \end{equation}
where $\Delta_{\pm}={\cal C}_9^{\rm eff}+ 2 \frac{m_b M_B}{s} ( {C_7^{\rm eff}}  \pm {C_7^{\rm eff}}^\prime       )$. The strong sensitivity to the coefficient   ${C_7^{\rm eff}}^\prime$   of the electromagnetic chirally flipped operator and an important enhancement factor $2 m_b M_B/s$ around 1 ${\rm GeV}^2$ are evident. 
\item The comparison between $A_T^{(2)}$ and $A_{FB}$ is particularly interesting: 
i) While $A_T^{(2)}$ is extremely sensitive to right-handed currents via  $C_7^\prime$ (and its CP violating phase), $A_{FB}$ shows only a very mild (for the modulus) or null (for the phase) sensitivity
 (see also Fig. 1, right and center plot). 
ii) Both observables exhibit a zero, or a lack of it, at the same value of $q^2_0$  at LO (but also at NLO)  if ${C_7^{\rm eff}}^\prime \neq 0$. iii) While $A_{FB}$ is only protected from large soft form factor uncertainties at its zero, $A_T^{(2)}$ is protected in the whole $1<q^2<6\, {\rm GeV}^2$ region. 
\item $A_T^{(2)}$ also serves as an excellent probe for a nontrivial  $C_{10}^\prime$. The latter  implies   a completely different $q^2$ dependence than a non-zero  coefficient  $C_7^\prime$.  
\item  $A_T^{(2)}$    can be measured using the one-angle projected angular distribution in the first run of data taking with the LHCb experiment and using the full angular distribution afterwards. See\cite{Egede:2008uy,Egede:2010zc} for a discussion of its experimental sensitivity.
\end{itemize}

The transverse observable   $A_T^{(5)}$, complementary to $A_T^{(2)}$,  was proposed  in 
Ref. \cite{Egede:2010zc}:
\begin{equation}
 \label{eq:AT5general}
 A_T^{(5)} = \frac{|A_{\|}^{R*} A_{\perp}^L+A_{\|}^{L} A_{\perp}^{R*}|}{|A_{\|}|^2+|A_{\perp}|^2}\,. 
\end{equation}
\begin{itemize}
 \item Contrary to $A_T^{(2)}$, $A_T^{(5)}$ exhibits a combination of left-right and $\perp$-$\|$ amplitudes that cannot be found in any single coefficient of the distribution. Its expression in terms of the coefficients of the distribution can be found using the explicit solution described in Sec.2: 
 \begin{equation}
A_T^{(5)} \Big\vert_{m_{\ell}=0}= \frac{\sqrt{16 J_{1}^{s\,2}-9 J_{6}^{s \,2} - 36 (J_{3}^2+J_{9}^2)}}{8 J_{1}^{s}} \,.  \nonumber 
\end{equation}
 
\item  In the large recoil limit and assuming  a nontrivial $C_{10}^\prime$,   $A_T^{(5)}$ simplifies to
 \begin{equation}
\label{eq:AT5SMLEET}
  A_T^{(5)}\Big\vert_{10^\prime} \!\!= \frac{\left| -C_{10}^{2} + |C_{10}^{\prime }|^2 + \left(2 m_b M_B C_7^{\rm eff}/q^2+ C_9^{\rm eff}\right)^2 \right|}
  {2 \left[C_{10}^{2} + |C_{10}^{\prime}|^2 + \left(2 m_b M_B C_7^{\rm eff}/q^2+ C_9^{\rm eff}\right)^2\right]}\  \,. 
\end{equation}
It implies that $A_T^{(5)}$ has a maximum value in the SM  ($C_{10}^\prime=0$)  of $1/2$ near to the position  of the zero of $A_{FB}$. 
 If $C_{10}^\prime \neq 0$ and $C_{10}^\prime < C_{10}$ the size of the local maximum decreases and its  distance to the SM maximum  is given by  $|C_{10}^{\prime NP}|^2/(C_{10}^2+|C_{10}^{\prime NP}|^2)$. This distance  can be used as a measurement of $C_{10}^\prime$ if  $C_{10}^\prime$ 
 represents the only contribution beyond the SM (see Fig.1, left plot). 
  \item Finally,  the position $q_0^2$ of the  maximum  moves if  $C_7^{\rm eff}$ or $C_9^{\rm eff}$ receives NP contributions like $A_{FB}$ (see again Fig.1, left plot). 
\end{itemize}

\section{Conclusions}

We have presented in detail a method  to construct observables, using the $K^*$ spin amplitudes as building blocks,  with high new physics sensitivity and reduced hadronic pollution. It is sufficiently general to be applied to other angular decays  with similar properties. The symmetries of the four-body decay, that play a central role in this method, are identified and interpreted. Finally, two observables are constructed fulfilling all the steps of the method and their properties are analyzed. 
$A_T^2$ emerges as an improved version of $A_{FB}$, containing almost all  the physical information of it but in a less QCD polluted way, and it also exhibits a much larger sensitivity to right-handed currents than $A_{FB}$.



\end{document}